\documentclass[aps,prd,twocolumn,superscriptaddress,a4paper]{revtex4}

\usepackage{graphicx}
\usepackage{eurosym}

\begin{document}


\title{Reply to arxiv:1106.3559 by J.I. Collar}



\author{The SIMPLE Collaboration}
\email{criodets@cii.fc.ul.pt}





\maketitle


Recent broadband comments by J. I. Collar \cite{arx1} have not only questioned the credibility of SIMPLE's recent Phase II results \cite{arx2,prl}, but impugned the competence of its collaborators. While normally we would not deign to make response, recent indications that these criticisms could actually be taken seriously force us to do so.

1. \textit{"limited SDD lifetime"}: the severest criticism of \cite{arx1} deals with concerns regarding the lifetime of the SIMPLE detectors, based on a "rapid aging" of the material leading to fractures, diffusion of the superheated liquid, pressure increases and gas leaks as per the author's "previous experience with SDDs".

Given the previousness of this experience, the real problem is what is meant by "rapid". The 20 d limit stated by the author (apparently determined only from the exposure and active mass of Fig. 3 of the cited Ref. 7) curiously differs from "present SIMPLE modules are stable over $\sim$ 40 d of continuous exposure" stated in the same Ref. 7, itself consistent with his also cited Refs. 4 and 6 (our \cite{jptese}). What is not however stated in \cite{arx1}, despite the frequent mention/use of PVP in its references, is that the limit was conservatively adopted for SDDs \textit{without} PVP in the fabrication, on the basis of signal avalanches which began to appear in the detectors after 40 days, essentially due to fractures and their propagation which the instrumentation of 2000 was unable to discriminate from bubble nucleations.

The aging concerns, which originated during the author's time in SIMPLE, are indeed "acknowledged" in \cite{arx2,prl}: there has seemed no particular need to continually re-"emphasize" the origins, given their extensive discussion over the past decade, to include not only the cited Refs. 4,5,7 of \cite{arx1} (which are also SIMPLE), but also their conveniently uncited confrontations \cite{astrop,tmtese,migfab,plb}. As noted in Ref. 7 of \cite{arx1}, much less in \cite{jptese}, the "rapid aging" process is slowed by the addition of PVP which increases the required gel fracturing energy and viscosity, strengthening the gel matrix while further reducing the already low solubility of the R-115. The studies cited in Ref. 5 of \cite{arx1} only report a part of the work detailed in \cite{tmtese} on the increase of SDD lifetime and fracture reduction via use of PVP of various polimerization indices, and other additives (largely post-dating the experience of \cite{arx1}). These studies included numerous neutron irradiations, which later in 2004-2007 were extended to include weak $\gamma$ sources in the development of gel recipes for long-lived SDDs based on refrigerants other than R-115 \cite{astrop}; while not directly examining the SDD longevity, these explored the rate of fracture occurrence with recipe variations.

Furthermore, the SDD lifetime is naturally increased if the device is only weakly irradiated \cite{jptese,tmtese}, i.e. the number of bubbles which can grow into fractures is small (it is in fact for just this reason that we do not irradiate before or during the measurements, but rather monitor the state of the each SDD electronically). Prior to 2006, all SDD fabrications were made in a Paris laboratory, which engendered the overland transportation of the SDDs to the underground experimental site in a state of "suspended animation". Analysis of this transport, beginning in 2004, indicated deleterious effects on the fabrications, to include installation in less than pristine condition (ie. with fractures, bubbles), and the formation of clathrate hydrates \cite{plb} which provide surfaces for bubble formation when being warmed to the device operating temperature of 9$^{\circ}$C: in 2006, a 210 mwe underground clean room within the LSBB was constructed, and the SDD fabrications relocated  - which led to substantial subsequent changes in the detector performances, including the absence of clathrate hydrates. In December of 2006, a run of 109 d with a R-115 SDD fabricated with new chemistry and submerged to the center of a 1.5x1.0x1.0 m$^{3}$ waterpool at 1500 mwe, resulted in a loss of detection stability only in the last two weeks of the run, and an operational lifetime increase of $\sim$ 2x. Naturally, fractures resulted from the Oswald ripening of the bubbles, and ultimately led to the performance degradation suggested by \cite{arx1} - but at the indicated significantly reduced rate. Nonetheless, the study of SDD gel composition, response and lifetime issues obviously has not ceased \cite{migfab}.

In short, the recent SIMPLE results "assume" nothing, especially as regards "perfect detector stability": we quote uncertainties in both temperature and pressure, we monitor both continuously, we only use those data obtained with a pressure increase of $\leq$ 0.2 bar (in Stage 1, $\sim$ 70 d of data satisfied the requirement; in Stage 2, the pressure was allowed to rise, with only $\sim$ 45 day of the data used in the science analysis). Monitoring of these parameters, together with the changes in signal ouput (capable of observing the SDD deterioration indicators of \cite{arx1} - see below) of the SDDs with time, permits knowledge of the state of each device \textit{throughout} its operation.

Incidentally, the SDDs were \textit{of course} inspected at the end of their runs, and showed no increased transparency nor change in gel coloration... thus failing the suggested "active volume depletion" test (which any competent experimentalist would naturally effect). Moreover, following each run, the R-115 was observed to evolve from the detectors upon warming, accompanied by the characteristic crackling "popcorn" sound of bubble nucleations.

2. \textit{"Instrumentation"}: the criticisms here are based on "marked differences" between the run and calibration neutron events, suggesting the former to be in fact "environmental acoustic noise" for which \cite{arx1} sees no "unambiguous" discrimination criteria.

Again, the problem returns to the author's "previous experience with SDDs", the change in instrumentation of which he appears not to appreciate. In 2006, the 2000 instrumentation  was replaced with an improved system \cite{migle2} providing a factor 100 noise reduction, and discrimination between various types of acoustic events associated with the SDD gel (fractures, trapped gas, and microleaks); in 2008, this was replaced by a fully microphonic response to all acoustic signal \cite{felizardo}, whether particle-induced, gel-associated or background noise; this system is based on a true high quality electret microphone as opposed to the piezo "buzzer" of 2000, and provides a factor 50 reduction in the previous noise levels, a voltage resolution of 0.3 mV, a timing resolution of 1.6$\times$10$^{-2}$ ms, and frequency resolution of 0.01 Hz. As documented, event analysis is based on examination of 3 signal parameters plus the individual power spectral densities of each; the cited Fig. 3a of \cite{mfidm}, possibly not well-resolved because of the Fig. compression, represents only a part of the parameter analysis as illustrative of the observed population separations -- the analysis is 3-dimensional, with the power spectral density a deciding final criterion, and is shown in \cite{felizardo} capable of clearly distinguishing between events originating from true nucleations, fractures, trapped N$_{2}$ gas, and microleaks. In fact, a library of acoustic calibration templates (including environmental acoustic noises) exists from which the acoustic origin of each event can be determined unambiguously, and particle-induced events identified with better than 97\% efficiency at a 95\% confidence level \cite{arx2,prl}.

Let's put this into context more concretely: in Stage 1 for example \cite{prl}, there were a total of 4056 signals recorded, of which 1828 were uncorrelated single events. Analysis of the 3 signal parameters plus the power spectral density in each case identified 88\% with various environmental acoustic noise events, 3.4\% in trapped N$_{2}$ gas, 0.11\% in N$_{2}$ escape, and 4.4\% in fractures. With the exposure reduction in \cite{arx2}, only 36 fracture events total are recorded over the $\sim$ 70 days of the 15 detector operation, or an average of 2.4 fractures per detector. More importantly perhaps, only 15 fractures were recorded by day 56 of the run, with only zero or one per SDD except in the case of two devices... somewhat in tension with the comment of \cite{arx1} that "SIMPLE modules feature only marginal improvements... ".

Note also that the ability of the current electronics/signal analysis to distinguish between bubble nucleations and gel fractures has serious impact on the "rapid aging" criticisms above, since the discrimination capacity permits operating a SDD with more than the 2 fractures responsible for the earlier 40 d lifetime limit \cite{jptese}, as well as monitoring its degradation during operation.

Thus said, Fig. 1 of \cite{prl} shows only a display of amplitude vs. frequency for the already-identified particle-induced event subset of both calibration and run events, where the "marked difference" lies only in the amplitudes. The background and calibration neutron distributions \textit{are} of course different: the background distribution is heavily moderated, whereas the offline neutron calibrations were naturally made with a weak Am/Be source moderated by only 15 cm of water in order to enhance the lower rate distribution tails, which would otherwise require more measurement time. None of the "neutron-like" events shown in \cite{prl} satisfy the criteria for any of the myriad "environmental acoustic noise" events, all of which lie outside the particle-induced frequency range and otherwise possess distinctly different signal parameters and power spectral densities.

The single neutron-like event identified in \cite{arx2} is "alleged" to have a neutron origin based on a reanalysis of its pulse shape which uncovered the overlapped double signature separation - \textit{both} of which satisfy all neutron-like event cuts. The comment in \cite{arx1} regarding the time separation between the two signatures is admittedly well-taken: our "$\sim$ 30 ms" should have been "$\sim$ 30 $\mu$s" (37 $\pm$ 8 $\mu$s, to be precise -- but which derives from the delay in sound propagation from the two nucleation events at different distances from the microphone, rather than the suggested "straggling of fast neutrons...in hydrogenated materials" which is of order ns), and we regret our lapse in having needlessly provoked this concern.

3. \textit{Simulation use}: although not based on "an actual measurement of the neutron flux and energy spectrum
at the experimental site", our simulations \textit{are} based on \textit{multiple} measurements of the U and Th impurities of all materials in the site shielding \cite{af2}, as well as the experiment geometry \& construction. The simulations, conducted in the blind with respect to the signal analysis, have yielded agreement with run results to within estimate uncertainties, in the process generating additional radio-assays and construction checks towards their self-improvement. Their initial intent was as a tool towards reducing the on-detector neutron background in the measurements; only later were they used in the results analysis when it became clear that the neutron background had not been entirely eliminated.

In the case of Stage 1, given the non-zero events observed (which the simulations suggested to be neutrons, but which could not be confirmed), we conservatively employed a well-accepted approach \cite{feldman} to this background subtraction, which inherently incorporates the uncertainties of the radioassays from which the background estimate is obtained. While there is of course no background subtraction in Stage 2, since the single neutron-like event could be identified as of neutron origin, it is again employed in the merging of the results. As we however conclude, while the merging might be questioned by the author and others, especially in view of its implied impact in the spin-independent sector of WIMP interactions, the Stage 2 result alone is sufficiently provocative to at least motivate a larger exposure measurement with further neutron background reduction.

In summary, although the concerns voiced in \cite{arx1} are important considerations in the use of SDDs in general, we find them based on extremely dated "previous experience with SDDs" which ignores the state-of-the-art of SIMPLE. The "known limitations" of the early days have become better understood and been addressed to a significant extent in both fabrication and instrumentation, yielding a substantially improved detector relative to that of 2000. In short, our SDDs are simply not the author's previous SDDs, and his concerns are unfounded as regards both the current SIMPLE program and its results -- which we continue to stand by as we have reported.


\end{document}